\documentclass[letterpaper]{article}

\usepackage{siunitx}
\usepackage{verbatim}
\usepackage{subcaption}
\usepackage{xcolor}

\DeclareSIUnit{\molar}{M}
\DeclareSIUnit{\gys}{Gy/s}

\usepackage[T1]{fontenc}

\usepackage{geometry}
\usepackage{setspace}

\usepackage[style = chem-acs]{biblatex}
\usepackage{graphicx}
\usepackage{float}
\newfloat{scheme}{htbp}{los}
\floatname{scheme}{Scheme}
\floatname{chart}{Chart}
\newfloat{graph}{htbp}{loh}

\usepackage{chemformula} 
\usepackage[version = 4]{mhchem} 

\usepackage{authblk}

\author[1,2]{Niels C. Gießelmann*}
\author[3]{Yasothaa Thavayogarajah}
\author[1,2,4]{Tobias Eklund}
\author[1,2]{Louisa E. Kraft}
\author[1,2]{Isabell Zick}
\author[2]{Yizhi Liu}
\author[1,2]{Fiona Berner}
\author[5]{Felix Lehmkühler}
\author[5]{Fabian Westermeier}
\author[3]{Mostafa Ahmadi}
\author[3]{Sebastian Seiffert}
\author[1,2]{Katrin Amann-Winkel**}

\affil[1]{Institute of Physics, Johannes Gutenberg University Mainz, Germany}
\affil[2]{Max Planck Institute for Polymer Research, Mainz, Germany}
\affil[3]{Institute of Chemistry, Johannes Gutenberg University Mainz, Germany}
\affil[4]{European XFEL, Schenefeld, Germany}
\affil[5]{Deutsches Elektronen-Synchrotron DESY, Hamburg, Germany}

\title{X-ray induced structural and dynamical changes in supramolecular polymer networks}

\date{*E-mail: niels.giesselmann@uni-mainz.de \\
**E-mail: amannk@mpip-mainz.mpg.de}

\begin{document}

\maketitle

\begin{abstract}
Beam-induced effects are key considerations in X-ray scattering studies of soft matter. Because these experiments typically aim to probe the unperturbed system, the structural and dynamic changes induced by the beam often remain underexplored. In the study at hand, the influence of X-rays on metallo-supramolecular tetra-arm poly(ethylene glycol) (tetra-PEG) networks is studied in dependence of the dose rate. Based on their transient interlinkage, such materials have the ability to reconnect and self-heal mechanical damage. Here, wide- and small-angle X-ray Scattering were employed to understand the effect of X-rays on the micro- and mesoscale structure of such polymer networks. Dynamical effects were probed using X-ray photon correlation spectroscopy. A restructuring of the polymer chains in the network is observed over several length scales, indicating a pre-crystallization step induced by the ionizing radiation, accompanied by expulsion of water from the network. In conjunction, a complex dynamical behavior is observed, consisting of an initial fast process, followed by an overall slow-down during the restructuring process. The two-time correlation functions exhibit an oscillatory behavior related to the development of multiple dynamic populations in the polymer network.

\end{abstract}



\section{Introduction}

Polymers and supramolecular networks with high water-content are ubiquitous in synthetic materials as well as in biological environments. Their collective molecular interactions give rise to a rich diversity of structural, mechanical, and dynamic properties. In particular, networks assembled by reversible non-covalent interactions provide versatile platforms for understanding and controlling the relationship between molecular organization, material properties, and emergent behavior. Supramolecular polymer networks consist of non-covalently cross-linked macromolecules.\cite{jangizehi2020defects} Such transient links can be established by functionalized polymer building blocks. A particularly well-defined material platform to conduct such investigations are supramolecular networks formed from uniform star-shaped poly(ethylene glycol) (PEG) building blocks with heterocomplementary interconnection,\cite{ahmadi2025reconfigurable, thavayogarajah2026metallo} extending the tetra-PEG approach by Sakai and co-workers for model-network formation to transient interlinkage.\cite{sakai2008design} This can be achieved, for example, through the use of building blocks with ligands that complex to metal ions, bringing together slim and weak with strong and sterically demanding ligands. This combination favors a heterocomplementary connection to reduce the amount of primary loop formation.

X-ray scattering based experimental methods provide a range of useful tools to better understand the structure and dynamical behavior of such systems. However, radiation damage is an important issue when conducting experiments using X-rays, especially on soft matter. During X-ray scattering experiments radiation damage can manifest in various ways, such as heating due to X-ray absorption or the formation of radicals due to ionization.\cite{ashfaq2020polymerization} These effects can lead to structural changes in the investigated materials and are therefore important to account for.\cite{susich1959radiation, bras2021x, jousseaume2023beam} The problem is well know for a long time, in literature one can find \textit{e.g.} reports going back to 1959 of X-ray induced crystallization of sucrose \cite{susich1959radiation}, as well as recent reports on long-range ordering of peptides \cite{cui2010spontaneous} or X-ray induced structuring of nano- and gel particles, to name only a few.\cite{takinami2023ionizing, riekel2025synchrotron} It has also been shown that the exposure to ionizing radiation can influence the mechanical properties of polymer systems, an effect that can also be made use of, for example, in radiation sensors.\cite{zhang2022recent}

Here, we investigate the influence of X-ray dose and dose rate on the micro- and mesoscopic structure using small- and wide-angle X-ray scattering (SAXS/WAXS), as well as the dynamical behavior of tetra-PEG based polymer networks based on heteroleptic metal-ligand complexes. Using X-ray photon correlation spectroscopy (XPCS), we see multicomponent dynamics evolving differently depending on the actual dose rate.


\section{Experimental}

\paragraph{Polymer network samples}

The preparation of heterocomplementary metallo-supramolecular hydrogels was based on two ligand-functionalized four-arm poly(ethylene glycol) (tetra-PEG) precursors, tetraDMPhen20k and tetraPhen20k, with a molecular weight of $\SI{20000}{\gram\per\mol}$ each. The tetraDMPhen20k precursor carries sterically hindering 2,9-bis(mesitylene)-1,10-phenanthroline (DMPhen) ligands, whereas tetraPhen20k is functionalized with 1,10-
phenanthroline (Phen) ligands. Copper(I)bromide (CuBr) was used to form coordination cross-links between the ligand-functionalized polymer arms. 

Following the established procedure, $\SI{75}{\micro \liter}$ of each polymer precursor stock solution was combined with $\SI{50}{\micro \litre}$ of a CuBr stock solution in acetonitrile, giving a sample volume of $\SI{200}{\micro \liter}$. The final ligand end-group and $\text{Cu}^+$ concentrations were $\SI{8}{\milli \molar}$ and $\SI{4}{\milli \molar}$, respectively, corresponding to a metal-to-ligand end-group ratio of $1:2$.

After mixing, acetonitrile was removed under vacuum at approximately $\SI{60}{\celsius}$ to obtain the dried coordination-cross-linked network. After cooling to room temperature, the network was rehydrated with $\SI{200}{\micro \litre}$ of milli-Q water. The samples were sealed and shaken for 24 h to facilitate swelling in water, yielding hydrogels with a final polymer concentration of $\SI{40}{\gram\per\liter}$.

This heterocomplementary approach incorporates both DMPhen- and Phen-functionalized tetra-PEG precursors into the same $\text{Cu}^+$-cross-linked network. A detailed description of the synthesis and characterization of the ligand-functionalized polymer precursors and the hydrogel preparation procedures is provided in our previously published work.\cite{thavayogarajah2026metallo}

\paragraph{X-ray scattering experiments}

\begin{figure*}[t]
    \centering
    \includegraphics[width=\linewidth]{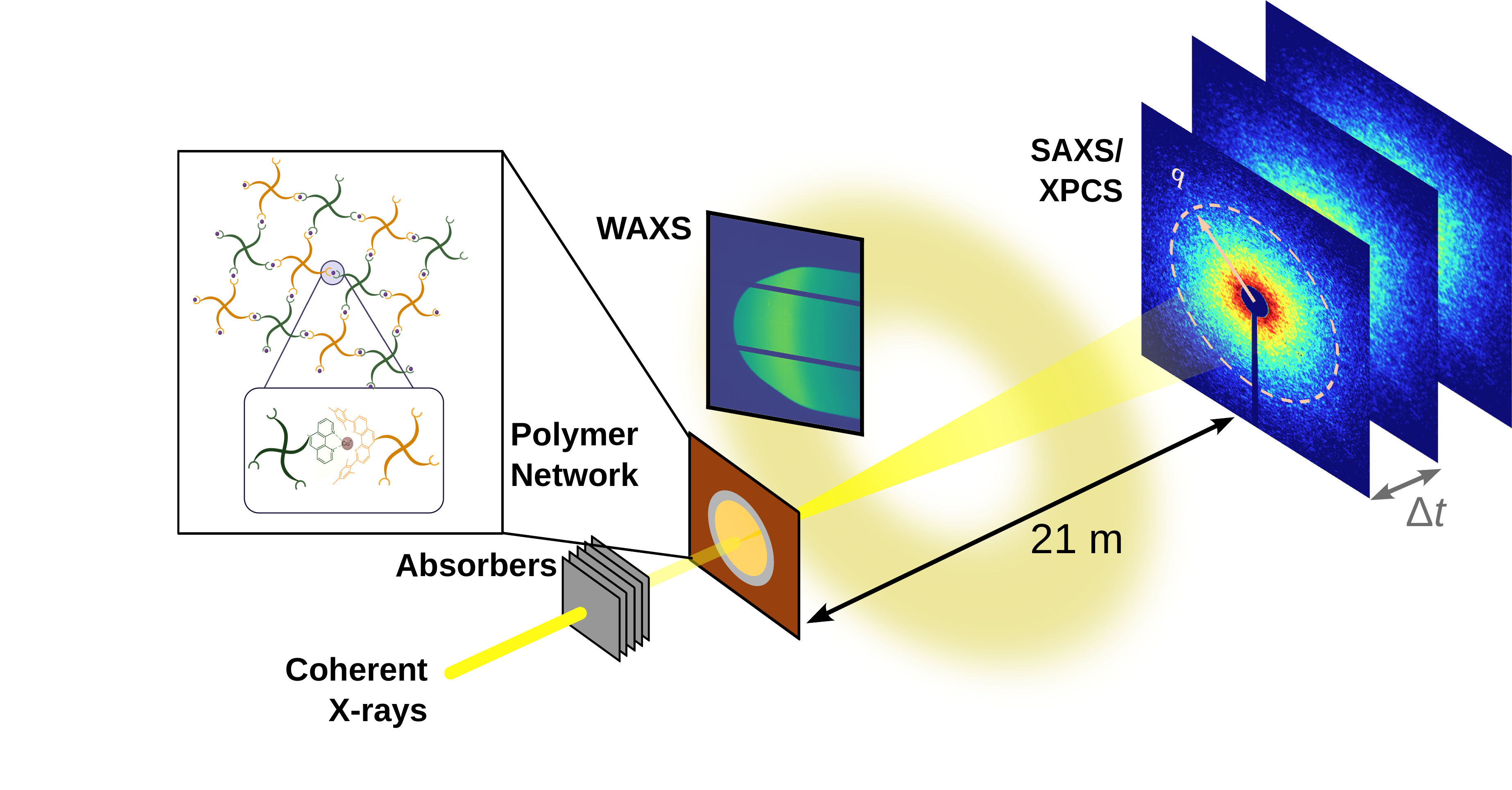}
    \caption{A sketch of the experimental setup. The coherent X-ray beam intensity is modulated by silicon absorbers before interacting with the sample. The scattered beam is detected in WAXS and in SAXS. The detector for the latter is positioned \SI{21}{\metre} downstream from the sample and records scattering patterns with a frame spacing of $\Delta t$, which are used for XPCS analysis. Measurements were done at different positions in the sample.}
    \label{fig:setup}
\end{figure*}

Time-resolved small- and wide-angle X-ray scattering (SAXS/WAXS) measurements were conducted at the Coherence Applications beamline P10 of the PETRA III facility at DESY in Hamburg (Germany) to study both the micro- and mesoscale properties of the polymer networks. A beam of $200 \times$\SI{200}{\micro\metre} was employed at an energy of \SI{8.05}{\kilo\eV}. While the WAXS geometry was used to monitor the microscopic structure of the network, SAXS was used to investigate the dynamics of the polymer network on macromolecular lengthscales.\cite{perakis2017diffusive, ladd-parada2022coherent, karina2025multicomponent}
In addition, we performed X-ray photon correlation spectroscopy (XPCS) experiments in SAXS geometry, details can be found below.

Fig.~\ref{fig:setup} shows a sketch of the experimental setup. The sample was placed in a dedicated holder in between two polyimide film windows at a distance of \SI{21}{\metre} upstream from the SAXS detector (\textit{DECTRIS EIGER X 4M}). To monitor the molecular structure of the polymer networks, WAXS was simultaneously conducted with a second detector (\textit{DECTRIS PILATUS}). 
In order to control the X-ray intensity on the sample, silicon absorbers were used to attenuate the incoming X-ray beam and the sample chamber. Each of the \SI{25}{\micro\metre} thick silicon wafers transmits \SI{70}{\percent} of the incident beam at the used photon energy, the transmitted fraction after $N$ absorbers can be determined as 

\begin{equation}
    f_{\text{tr.}} = \frac{I_{\text{tr.}}}{I_0} = 0.7^{N}.
\end{equation}

\noindent We can furthermore approximate the absorbed X-ray dose rate $\dot{D}$ during the measurement as \cite{jeffries2015limiting}.

\begin{equation}
    \dot{D} = 1000 \cdot \frac{Etf}{\rho_{\text{m}} A L} \left[ 1 - \frac{1}{\exp\left({\frac{\mu}{\rho} \rho_{\text{m}} L} \right)} \right]
\end{equation}

\noindent where $E$ is the X-ray energy in Joule, which, at a photon energy of \SI{8.05}{\kilo\eV} is \SI{1.13e-15}{\joule}. $f=f_{\text{tr.}}\cdot f_{\text{initial}}$ is the photon flux on the sample with the initial flux $f_{\text{initial}}$ being \SI{1e11}{\per\second}. $t$ is the exposure time, $L=$\SI{0.1}{\centi\metre} the sample thickness and $A$=\SI{1e-4}{\cm\squared} the exposed beam area. $\mu/\rho$ is the attenuation cross section of the sample and was calculated using the \textit{XCOM} software by NIST \cite{XCOM} to be \SI{8.245}{\centi\metre\squared\per\gram}. The mass density of the sample $\rho_{\text{m}}$ was approximated by that of water as the sample consists of \SI{96}{\percent} water by weight. 

The exposure time $t$ and the number of absorbers are the only parameters which are varied throughout the experiment, allowing us to study the flux-dependence of the structure and dynamics of the polymer network. By using different amounts of absorbers and varying exposure times, it is therefore possible to isolate the influence of beam-induced effects on the these properties of the sample.
The different number of absorbers used, as well as the resulting transmissions and absorbed doses, are listed in Table~\ref{tab:absorbers}.

After each measurement, the position in the sample was shifted to ensure an unexposed spot for the next measurement. WAXS data reduction was conducted using the \textit{pyFAI} python package \cite{ashiotis2015pyfai,kieffer2020pyfai}.

\begin{table*}[h]
    \centering
    \begin{tabular}{|r|r|c|}
    \hline
 \# Absorbers & Transmission [\%] & Dose rate $\dot{D}$ [Gy/s]  \\ \hline
        16 & 0.3 &  $24 \pm 4$   \\
        12 & 1.3 & $98\pm 15$\\
        8  & 5.8 & $409 \pm 61  $\\
        6 & 11.8 & $835 \pm 124 $\\
        4  & 24.0 & $1704 \pm 253 $\\ \hline

    \end{tabular}
    \caption{Number of absorbers (\SI{25}{\micro\metre} thick Si), calculated X-ray transmission and dose rate $\dot{D}$.}
    \label{tab:absorbers}
\end{table*}

In order to investigate the polymer network dynamics, X-ray photon correlation spectroscopy (XPCS) was employed, making use of the coherent properties of the \SI{8.05}{\kilo\eV} X-ray beam of P10. In this method, the intensities of speckles at a certain $q$ on the detector are correlated over time. \cite{lehmkuhler2021femtoseconds} For any two times, this can be described as

\begin{equation}
    C(q, t_1, t_2) = \frac{\langle{I(q,t_1)} I(q, t_2) \rangle}{\langle I(q, t_1) \rangle  \langle I(q, t_2) \rangle},
    \label{eq:two-time}
\end{equation}

\noindent where $t_1$ and $t_2$ are the two times and $q$ the absolute value of the scattering vector, which is defined as 

\begin{equation}
    q = \frac{4 \pi}{\lambda} \sin{\left(\frac{2\theta}{2} \right)}
\end{equation}

\noindent with the X-ray wavelength $\lambda$ and the scattering angle $2\theta$.

In a stable system, which does not evolve over time, the resulting matrix will show a uniform shape with a maximum on the diagonal. A matrix, where the shape of the diagonal changes over time indicates a non-equilibrium system that is subject to change during the measurement period. The two-time correlation function can therefore be used as a measure of dynamic heterogeneity of the sample during the measurement period.\cite{lehmkuhler2020emergence, lehmkuhler2021femtoseconds, begam2021kinetics, anthuparambil2023exploring, karina2025multicomponent}

To extract sample relaxation throughout the measuring process, $g_2$-like one-dimensional sections of the two-time correlation function $C(t_{\text{w}}, t_{\text{w}}+\Delta t)$ can be extracted, starting at a certain wait time $t_{\text{w}}$.\cite{lehmkuhler2020emergence, lehmkuhler2021femtoseconds, anthuparambil2023exploring} This way, changes in the dynamic behavior during the measurement can be quantified, as these can be fitted according to the Siegert relation \cite{lehmkuhler2021femtoseconds}

\begin{equation*}
    C(t_{\text{w}}, t_{\text{w}}+\Delta t) \equiv g_2(q,\Delta t) = 1 + \beta |f(q,\Delta t)|^{2}  
\end{equation*}

where $\beta$ is the scattering contrast and $f(q,\Delta t)$ the first order autocorrelation function. A single decay can typically be described by a stretched exponential function

\begin{equation*}
    f(q, \Delta t) = \exp \left(- \frac{\Delta t}{\tau} \right)^{\gamma}
\end{equation*}

In the present case, $f(q, \Delta t)$ can be approximated as two stretched exponential decay functions:

\begin{equation*}
    f(q, \Delta t) = A_1 \exp \left(\frac{\Delta t}{\tau_1} \right)^{\gamma_1} + A_2 \exp \left( \frac{\Delta t}{\tau_2} \right)^{\gamma_2}
\end{equation*}

Here, $A_1$ and $A_2$ are the amplitudes of the two decays, $\tau_1$ and $\tau_2$ the relaxation times and $\gamma_1$ and $\gamma_2$ the stretching exponents, also called the \textit{Kohlrausch-Williams-Watts} exponent.\cite{lehmkuhler2021femtoseconds}

The XPCS measurements were done in SAXS geometry to resolve dynamics on macromolecular and mesoscopic length scales. The processing of this data, including the computation of the two-time correlation functions $C(q,t_1,t_2)$ and the azimuthal integration in SAXS, were done using the XPCSGUI software, made available by beamline P10.
WAXS and SAXS measurements were taken simultaneously with 1 pattern per second and \SI{1}{\second} exposure time.


\section{Results and discussion}

\paragraph{Structural behavior}

\begin{figure*}
    \centering
    \includegraphics[width=\linewidth]{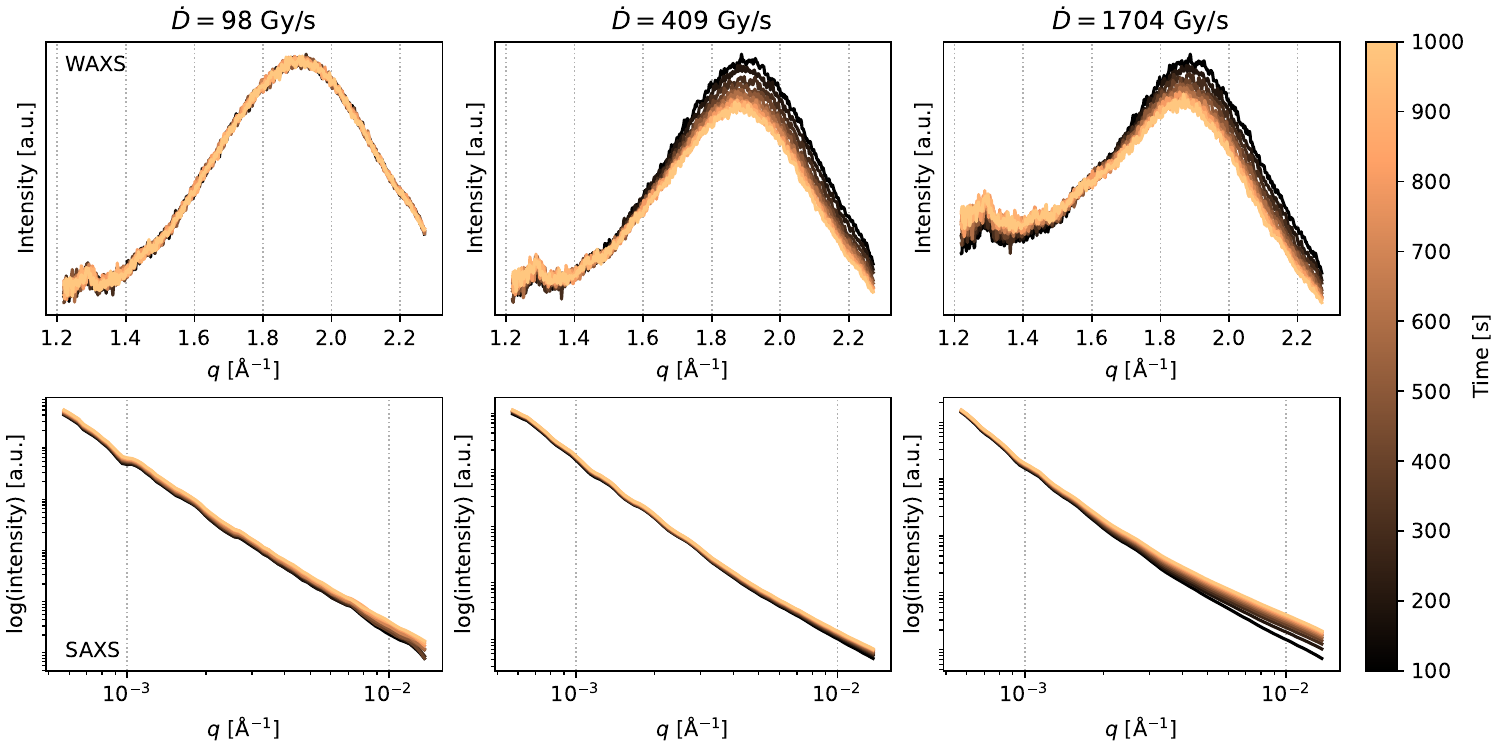}
    \caption{WAXS (\textbf{top}) and SAXS (\textbf{bottom}) of a tetraDMPhen/Phen 20k polymer network, measured with a dose rate of 98, 409 and \SI{1704}{\gys}. Each curve is averaged from 100 single patterns, each exposed for 1s.}
    \label{fig:waxs-saxs}
\end{figure*}

The top row of Fig.~\ref{fig:waxs-saxs} shows the WAXS intensity over time for different dose rates. Among the 1000 patterns taken in total,  100 each have been averaged. The brighter the color of the plot, the later in the measurement the patterns were taken. 

All plots show a clear maximum around $q\approx\SI{1.9}{\per\angstrom}$, the characteristic broad diffraction peak of water.\cite{perakis2017diffusive} For the lowest dose rate of \SI{98}{\gys}, a stable signal is observed with no changes over time. With increasing beam dose rate (409 and \SI{1704}{\gys}), the signal starts to change over time. The primary maximum decreases and narrows over time accompanied by a shift towards smaller $q$. Additionally, a general increase in the range of $\SI{1.2}{\per\angstrom} < q < \SI{1.6}{\per\angstrom}$ is observed. The small peak in this region, located at $q \approx \SI{1.3}{\per\angstrom}$ is associated to the polymer building blocks rather than water.

\begin{figure*}
    \centering
    \includegraphics[width=\linewidth]{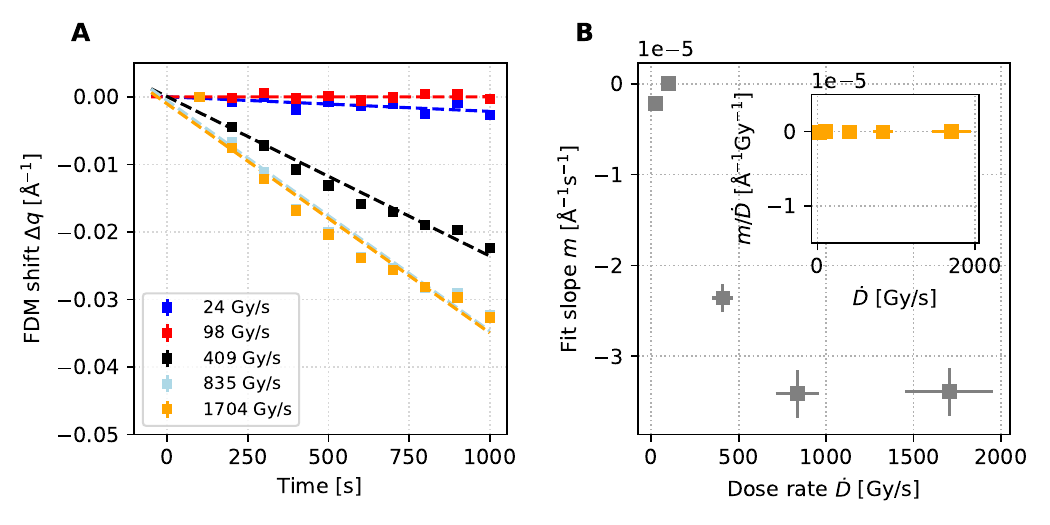}
    \caption{\textbf{A} The $q$-position of the first diffraction maxima (FDM) is determined from the WAXS data. Here the relative shift $\Delta q$ over the measurement time is shown. The dashed lines are linear fits. \textbf{B} The dose rate in Gray per second versus the values of the slope from the linear fits shown in \textbf{A}. The inset shows the value of the slope divided by the dose rate per second.}
    \label{fig:peak_shift}
\end{figure*}

Figure~\ref{fig:peak_shift}A shows the $q$-shift of the position of the main WAXS peak as a function of experimental time for the different attenuations. While the $q$ stays almost constant for a dose rate of \SI{24}{\gys} (blue), a decrease in intensity is observed for dose rates of 409 (black) and \SI{1704}{\gys} (yellow). In order to probe if this decrease is directly dependent on the dose rate, the first diffraction maximum (FDM) was fitted using a Lorentzian function in order to determine its $q$-position. The so-derived time-dependent peak shift is shown in Fig.~\ref{fig:peak_shift}A. These curves were fitted by linear functions as a first approximation. The slopes of the fit results in relation to the dose rate are shown in Fig.~\ref{fig:peak_shift}B, where a dependence of the slope on the dose rate is observed. If the values of the slopes are divided by the respective dose rate, this dependence disappears, as shown in the inset. This demonstrates that the $q$-shift observed in the WAXS maxima is directly correlated with the applied dose rate.

This is, the increase in the WAXS signal at $q < \SI{1.5}{\per\angstrom}$ hints at a re-structuring of the PEG in the system. It is know that crystalline PEG exhibits Bragg peaks at $q \approx \SI{1.35}{\per\angstrom}$ and $q \approx \SI{1.6}{\per\angstrom}$, which relate to the (210) and (032)* reflexes. \cite{golitsyn2019crystallization, takahashi1973structural} While we do not see an emergence of crystalline peaks here, an increasing intensity around this $q$-range indicates that some re-structuring of the PEG might be taking place, perhaps as a pre-crystallization step, where the PEG chains form structures with varying correlation lengths, thereby raising the overall signal in this $q$-range, rather than leading to distinctive peaks. 
As the characteristic water peak is decreasing, one can assume that, while the sample in general is not drying, water is still expelled from parts of the polymer network by this restructuring process and a local separation of water and polymer is taking place.

This interpretation is supported by the observed changes in the SAXS signals (Fig.~\ref{fig:waxs-saxs} (bottom-row)). When focusing on the large-$q$ region of the SAXS data ($q>\SI{5e-3}{\per\angstrom}$), especially at the highest dose rate of \SI{1704}{\gys}, a time-dependent intensity increase is observed. The signal in this $q$-region is dominated by the polymer network on a mesoscopic length scale. Intensity changes at this length scale support the interpretation from the WAXS data, namely a restructuring of the polymer network accompanied by water being expelled from the network. This way, the SAXS regime is becoming stronger due to a higher fraction of polymer in the sample, while the WAXS intensity decreases due to a decreasing amount of water.

One potential explanation for the re-structuring could be a heating effect caused by the X-rays. However, with the given experimental conditions, heating should not play a major role in a water based sample. Another possible reason could be drying of the sample. However, as the sample cell is sealed, no water should get out of the system and the sample did not show significant mechanical changes after the measurement. Therefore, the most likely source of the structural changes is ionization in the PEG chains, which in turn leads to the observed re-structuring.

\begin{figure*}
    \centering
    \includegraphics[width=\linewidth]{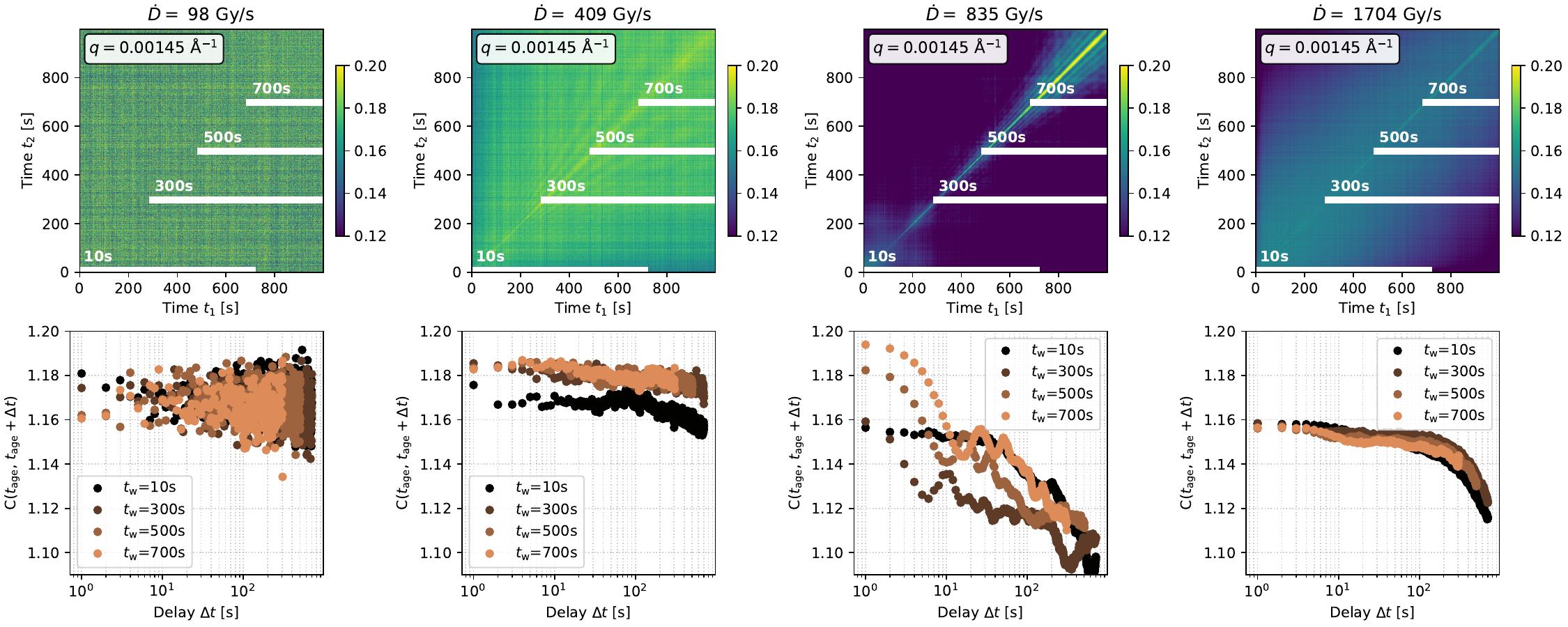}
    \caption{Two time correlation functions (TTC) measured at dose rates 1.3, 5.8, 11.8 and \SI{1704}{\gys} (top row). Correlation functions have been extracted by tracing along the white lines starting at \SI{10}{\second}, \SI{300}{\second}, \SI{500}{\second} and \SI{700}{\second}, respectively. The corresponding correlation functions $C(q,t_{age},t_{age}+\Delta t)$ are plotted in the bottom row.}
    \label{fig:TTCF}
\end{figure*}

\paragraph{Dynamical behavior}

The top row of Fig.~\ref{fig:TTCF} shows the two-time correlation (TTC) function matrices for 98, 409, 835 and \SI{1704}{\gys}. Extracted cuts after waiting times $t_{\text{w}}$ of 10, 300, 500 and \SI{700}{\second} are plotted below each two-time matrix. At \SI{98}{\gys} the matrix, as well as the extracted cuts, look flat with no visible decay. This indicates a stable sample with no observable dynamics on the measured time scale of \SI{1000}{\second}. Increasing the photon flux (\SI{409}{\gys}), we observe a change in the dynamical behavior of the polymer network. Here, an initial fast decay with a relaxation time in the range of seconds is observed. With increased waiting time, this decay slows down with oscillations in the correlation function emerging around \SI{100}{\second}, also visible as stripes in the TTC image. Increasing the dose rate further to \SI{835}{\gys}, the behavior throughout the measurement becomes even more heterogeneous, with a slow-down of the relaxation during the first \SI{100}{\second}. At this point, the sample undergoes a rather drastic change, with fast, oscillating features appearing in the two-time matrix, also well visible in the extracted horizontal cuts. This complex dynamics persists for the remaining measurement, while the overall dynamic slows down. To summarize, the TTC (\SI{835}{\gys}) first gets narrow (acceleration) and then the funnel opens up, indicating a complex decelaration in the dynamics. 

It is notable that the heterogeneous dynamical behavior observed  at 409, 835 and \SI{1704}{\gys} appears in combination with the structural changes observed in WAXS, as discussed above. It would be expected that a restructuring of the PEG network would lead to changes in the dynamics as well. The slow-down observed in the \SI{409}{\gys} and \SI{835}{\gys} TTC matrix could indicate such a transition. 

Due to the complexity of the matrices, it is generally difficult to identify individual processes that can be directly linked to the observed structural change. Oscillations, as observed in the matrices at higher $\dot{C}$ have been identified as deriving from heterodyne scattering components \cite{li2023intrinsic, karina2025multicomponent, He2024, He2025}, meaning that more than one population contributes to the scattering signal. If we assume that X-ray exposure leads to restructuring or pre-crystallization in the PEG chains. the different dynamic populations could represent building blocks that are at different points in this process.

The last matrix shows the two-time correlation for \SI{1704}{\gys}, where an initially fast decay is observed that slows down, developing oscillations at $t<\SI{200}{\second}$, similar to the measurement at \SI{409}{\gys} (see SI), before gradually speeding up again for the remainder of the measurement.

\begin{figure}[t]
    \centering
    \includegraphics[width=0.45\textwidth]{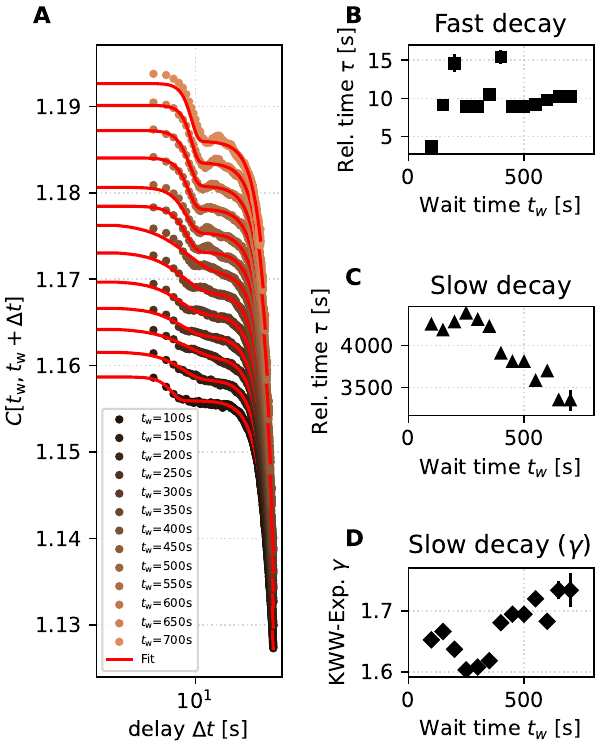}
    \caption{\textbf{A} Extracted correlation functions $C(q,t_\text{{w}},t_\text{{w}}+\Delta t)$, from horizontal cuts from the TTCF at \SI{1704}{\gys} including fits. Curves are shifted horizontally for clarity. Relaxation times for the fast and slow components of the double stretched exponential decay (\textbf{B} \& \textbf{C}), as well as the \textit{KWW}-Exponent $\gamma$ for the slow decay (\textbf{D)} over different waiting times $t_{\text{w}}$}.
    \label{fig:decay-fit}
\end{figure}

The dataset obtained with the highest dose rate was analyzed in more detail, which is  visualized in Fig.~\ref{fig:decay-fit}. We extracted horizontal cuts to the TTC at a constant delay of \SI{50}{\second} (\SI{1704}{\gys}). The so derived correlation functions have been fitted  with a double-exponential decays (Fig.~\ref{fig:decay-fit}A), similar to previous studies. \cite{ladd-parada2022coherent, karina2025multicomponent, eklund2026global}  As a evolution of time, one observes the correlation functions to clearly consist of two processes, a persistent slow decay, as well as a fast decay (B) that gets more prominent with increasing measuring time. The tail of this fast decay is already visible during the first \SI{100}{\second}, while it gets more prominent and slows down with increasing waiting time. Fig.~\ref{fig:decay-fit} B and C shows the extracted relaxation times of the two processes. Here, the effects observed before get clearer. A general increase is observed in the fast decay with an apparent step at around \SI{600}{\second}, with two outliers in the data point. Meanwhile the relaxation time of the slow decay decreases steadily from 4250 to about \SI{3300}{\second}. 

The \textit{KWW-}exponent of the fitted slow decay is larger than 1.6 for all extracted cuts, as shown in  Fig.~\ref{fig:decay-fit}D, an effect which could potentially be explained by stress relaxations in the system \cite{li2023intrinsic}. The re-structuring of PEG chains in connected building blocks in the network could cause excessive stress on the network. If the stress becomes too large, the connection might separate, leading to a dynamic with $\gamma>1$.

Overall, the here observed complex dynamics in measurements with higher dose rate, are consistent with the observation in WAXS and a potential pre-crystallization of the PEG arms in the polymer network. A general slowing down indicates an inhibition of the transient binding and a formation of more permanent structures. The relative stability of the dynamic behavior at \SI{1704}{\gys} after the initial slow-down indicates the transition to a different structural state compared to the initial transient polymer network. Furthermore, the transition of this change depends on the X-ray intensity, as shown by the presence of the oscillations throughout the entire measurements at \SI{409}{\gys} and its disappearance after approximately \SI{200}{\second} at \SI{1704}{\gys}

\newpage
\section*{Summary}

We have studied the influence of X-ray beam intensity on the structure and dynamics of tetra-PEG based metallo-supramolecular polymer networks in order to assess the nature and influence that radiation damage can have on such materials. Due to the transiently linked building blocks, these materials have the ability to reconnect or even self-heal. The influence of defects induced by x-ray radiation, however, remained unclear. We here used an initially stable and homogeneous system \cite{thavayogarajah2026metallo} as starting material, performing time-dependent and dose-rate dependent scattering experiments. In total, we used four different conditions with a dose rate from 24 to \SI{1704}{\gys}, which is calculated from the X-ray transmission through the Si-absorbers. It was observed that an increase in X-ray intensity leads to a re-structuring of the polymeric network, which can be seen as an overall increase in SAXS intensity, as well as an increasing intensity at around $\SI{1.2}{\per\angstrom} < q < \SI{1.6}{\per\angstrom}$ in the WAXS data. 
Simultaneously, a decrease in the primary diffraction maximum at $q \approx \SI{1.9}{\per\angstrom}$, fingerprint region for water, is hinting at an expulsion of water from the network.
Furthermore, XPCS was used to study the network dynamics. Here, the two-time correlation matrices (TTC) reveal the emergence of a complex dynamical behavior as a result of the X-ray radiation.
The network undergoes multiple dynamical transitions during the X-ray exposure. While the sample is stable at the lowest dose rate of \SI{98}{\gys} with no relaxation decay on the probed timescales of 1000 s. Higher dose rates induce heterogeneous dynamics, which slow down during the first 100 s and subsequently speed up again (see Figs.~\ref{fig:TTCF}). 
Taken together, the structural changes in SAXS and WAXS and the complex dynamical behavior observed through XPCS suggests a potential pre-crystallization step, originating from the ionizing effect of the absorbed X-rays. Such a state alters the molecular and mesoscale structure of the polymer network, while also slowing down the dynamics, which were previously induced into the system by elevated X-ray flux. Two-time correlation functions show emerging oscillations within a process of an overall dynamic slow-down. This could be explained by heterodyne scattering, caused by scatterer populations that are more or less strongly affected by the X-ray exposure or areas of static behaviour such as partly crystallized patches. Fitting the slow decay at the highest dose rate revealed \textit{KWW}-exponents $\gamma>1$, indicating stress relaxation is present in the damaged polymer network.

Although radiation-induced effects are a major consideration in X-ray scattering experiments on soft matter, their consequences for the material are complex and far from fully understood. Our results show that combining structural (SAXS/WAXS) and dynamic (XPCS) analysis approaches is essential to disentangle these effects, and that careful dose assessment should be an integral part of such experiments.

\section*{Data Availability}

The acquired SAXS and WAXS data, as well as the analysis code are freely available at [INSERT ZENODO LINK]

\section*{Acknowledgements}

The authors acknowledge funding from the \textit{Deutsche Forschungsgemeinschaft} (DFG, German Research Foundation): N.C.G., Y.T., T.E., L.E.K. and K.A.-W. acknowledge funding in the framework of the collaborative research center \textit{Defects and Defect Engineering in Soft Matter} (\textit{SFB1552}) under project number 465145163. K.A.-W. is a member of the RTG 2516 (Grant No. 405552959). I.Z. was the recipient of a doctoral position within the RTG 2516 program.
T.E. acknowledges funding from the Centre for Molecular Water Science (CMWS) within an Early Science Project.
We acknowledge DESY (Hamburg, Germany), a member of the Helmholtz Association HGF, for the provision of experimental facilities.
Parts of this research were carried out at PETRA III.
Data was collected using the P10 beamline operated by DESY Photon Science.
Beamtime was allocated for proposal I-20240452.
This research was supported in part by the Maxwell computational resources operated at DESY, Hamburg, Germany.

\section*{Supporting information}

The supporting information contains a magnified image of the

\printbibliography

\end{document}